\documentclass[12pt,twoside,a4paper]{article}
\usepackage[dvips]{epsfig}
\renewcommand{\thefootnote}{\fnsymbol{footnote}}

\def\gsim{\:\raisebox{-0.5ex}{$\stackrel{\textstyle>}{\sim}$}\:}

\begin{document}
\title{
\vskip-3cm
{\baselineskip14pt \normalsize
\hfill MZ-TH/00--46 \\ 
\hfill hep--ph/0011221 \\ 
\hfill November 2000 \\}
\vskip4cm
Hadronic Corrections at ${\cal O}(\alpha^2)$ to the \\[1ex]
Energy Spectrum of Muon Decay\footnote{Supported by the Deutsche
  Forschungsgemeinschaft.} \\[3ex]
\author{A.~I.~Davydychev\footnote{On leave from Institute for Nuclear
    Physics, Moscow State University,
    119899 Moscow, Russia. Email: davyd@thep.physik.uni-mainz.de}~, 
K.~Schilcher and 
H.~Spiesberger\footnote{Email: hspiesb@thep.physik.uni-mainz.de} \\[2ex]
\normalsize{Institut f\"ur Physik, Johannes-Gutenberg-Universit\"at,}\\ 
\normalsize{Staudinger Weg 7, D-55099 Mainz, Germany} \\[10ex]
} }
\date{}
\maketitle
\begin{abstract}
\medskip
\noindent
We consider the impact of ${\cal O}(\alpha^2)$ hadronic corrections to
the energy spectrum of the decay electron in muon decay. We find that
the correction can be described, within good approximation, by a linear
function in the electron energy. Explicit expressions for the form
factors needed in an approach based on dispersion integrals are given.
\end{abstract}
\thispagestyle{empty} 

\renewcommand{\thefootnote}{\arabic{footnote}}
\setcounter{footnote}{0}
\newpage 



\noindent
{\bf 1.} Testing the electroweak sector of the Standard Model requires
to fix the coupling constants of the Lagrangian by relating a number of
high-precision experiments to theoretical predictions. Ideally,
experimental data and theoretical predictions should be known with
comparable precision. Apart from the electromagnetic coupling constant
$\alpha$ and the $Z$-boson mass $m_{Z}$, the input of choice is the muon
life time $\tau_{\mu}$, with the present best value
$\tau_{\mu}=(2.19703\pm0.00004) \, \mu s$ \cite{PDG}.  Experiments are
planned at the Paul Scherrer Institute \cite{mulan} and the Rutherford
Appleton Laboratory \cite{riken} which would reduce the experimental
error on the muon life time by more than one order of magnitude.

These experimental data have reached such a precision that quantum
corrections can be observed.  To match this accuracy from the theory
side, two-loop radiative corrections to the muon decay in the full
electroweak Standard Model are needed. This is a formidable, but not
impossible, task. A step in this direction is the calculation of the
purely electromagnetic corrections to order $\alpha^{2}$ in the Fermi
theory, which has been performed by van Ritbergen and Stuart
\cite{Rstuart1,Rstuart2} (see also \cite{steinhauser}). Besides that,
also the ${\cal O}(N_f \alpha^2)$ corrections in the Standard Model have
been considered in Refs.\ \cite{Malde,Freitas}.

It is well known that the decoupling theorem is not applicable to the
life time calculation, which means that it is mandatory to also include
the contribution of the heavy degrees of freedom. In contrast, the
decoupling theorem is fully applicable to the {\em normalized} electron
energy spectrum in muon decay, which has also been measured with an
amazing accuracy at the per-mille level. Again the experimental error is
expected to be reduced by more than one order of magnitude in a future
experiment at TRIUMF \cite{triumf}. The spectrum calculation is
different from the one for the life time since the
Kinoshita-Lee-Nauenberg theorem \cite{KLN} is not in effect.
Consequently, powers of the large logarithm $\ln(m_{\mu}/m_{e})$ do not
cancel in the calculation of the electromagnetic corrections to the
spectrum. This becomes obvious when fitting the spectrum corrected to
order ${\cal O}(\alpha)$ to the Michel spectrum: the resulting effective
Michel parameter differs by about $6\,\%$ from its lowest-order value
\cite{drho}, a correction which is more than 10 times larger than the
corresponding correction to the muon life time. At order ${\cal
  O}(\alpha^{2})$ the radiative corrections may be expected to be of the
order of several per-mille, i.e.\ they could possibly be visible in
present data already, not to speak of future high-precision experiments.

Given this perspective we present in this note the calculation of the
hadronic contribution to the energy spectrum of the final-state electron
in muon decay. This contribution is not expected to be logarithmically
enhanced, but nonetheless is required for an eventual complete
second-order calculation. The details of the calculation are given in
the following sections.

\vspace{3mm}




\noindent
{\bf 2.}  We consider the decay of a muon in its rest system,  
\begin{equation}
\mu^-(p) \rightarrow e^-(p^{\prime}) + \nu_{\mu}(q_1) + \bar{\nu}_e(q_2)
\, , 
\label{mdecay}
\end{equation}
and define momenta as shown in (\ref{mdecay}).  The momentum transferred
from the charged particles to the neutrino pair, $q = q_1 + q_2$, is
then given by
\begin{equation}
q = p - p^{\prime} \, .
\end{equation}
It is convenient to introduce the dimensionless variable
\begin{equation}
x = \frac{2E_e}{m}
\end{equation} 
to denote the ratio of the energy of the decay electron $E_e$ with
respect to the muon mass $m$. Neglecting the electron mass,
$p^{\prime 2} = 0$, we see that the kinematically allowed range is
\begin{equation}
0 \le x \le 1 \, ,
\end{equation} 
and one has 
\begin{equation}
q^2 = (1 - x) m^2 \, .
\end{equation}

The matrix element ${\cal M}$ for (\ref{mdecay}) in the Fermi theory can
be calculated most conveniently after a Fierz rearrangement factorizing
the amplitude into a current $J_{\mu}$ which describes the $\mu e$
transition and a current for the $\nu_{\mu} \nu_e$ interaction. After
squaring and summing (averaging) over spins, one can write $|{\cal
  M}|^2$ as a product of two corresponding tensors. The one pertaining
to the neutrino interaction can be integrated over the unobserved
momenta of the neutrinos independently, leading to
\begin{equation}
N_{\mu \nu} = q_{\mu} q_{\nu} - g_{\mu \nu} q^2 \, .
\label{nmunu}
\end{equation}
This tensor will be contracted with 
\begin{equation}
C_{\mu \nu} = J^{\ast}_{\mu} J_{\nu}\, , 
\quad {\rm with} \quad 
J_{\mu} = \bar{u}_e(p^{\prime}) \Lambda_{\mu}(q) u_{\mu}(p) \, ,
\end{equation}
where $\Lambda_{\mu}(q)$ is the effective vertex of the four-fermion
interaction.  At the lowest order, $\Lambda_{\mu}(q)$ is identified with
\begin{equation}
\Lambda_{\mu}^0 = \frac{G_F}{\sqrt{2}} \gamma_{\mu} 
\left(1 - \gamma_5\right) \, ,
\end{equation}
where $G_F = (1.16637 \pm 0.00001) \times 10^{-5}$ GeV$^{-2}$ is the
Fermi coupling constant.

\begin{figure}[tb]
\unitlength 1mm
\vspace{-1cm}
\begin{picture}(120,53)
\put(5,0){\epsfig{file=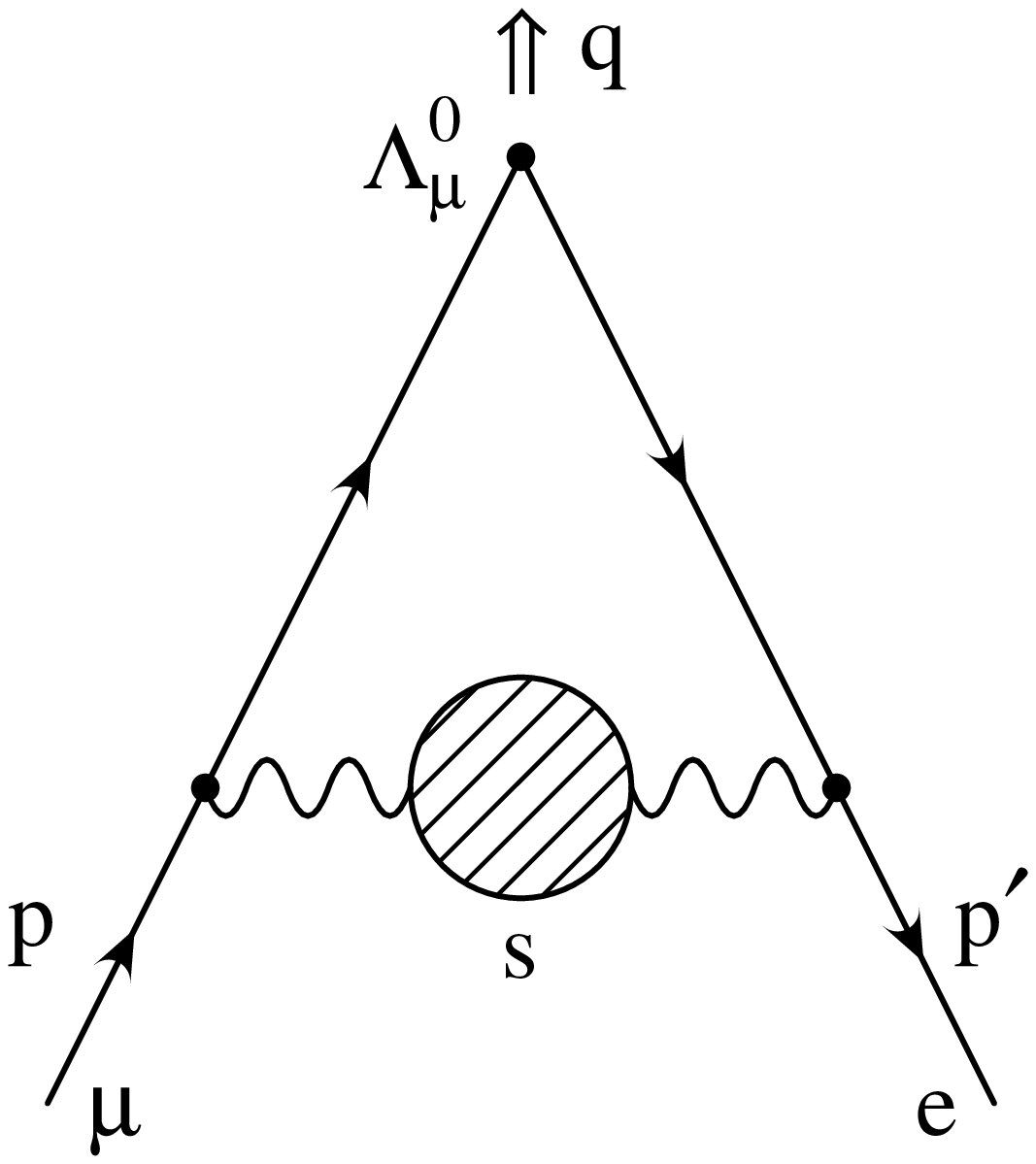,,width=4.97cm}}
\put(55,0){\epsfig{file=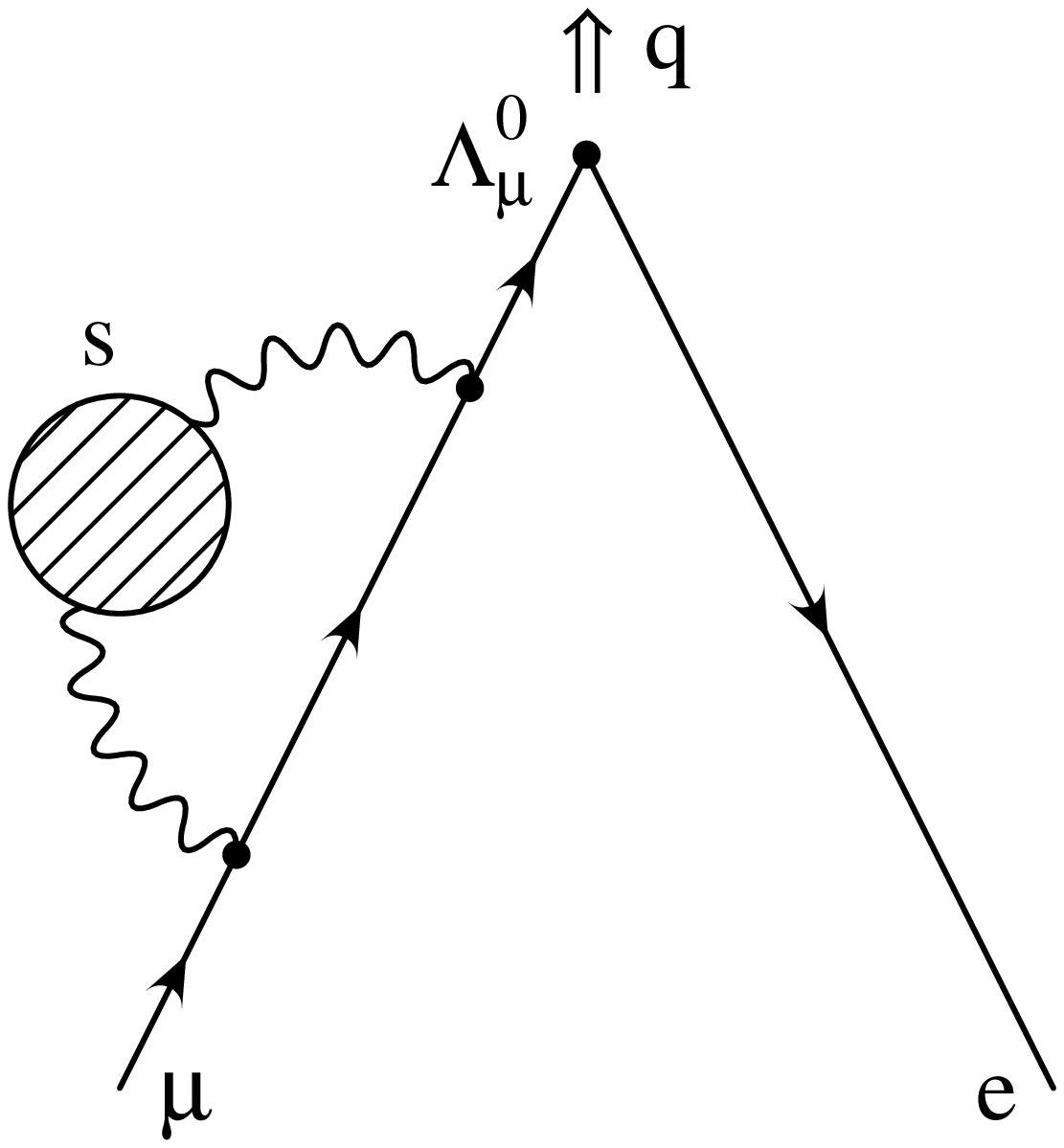,,width=4.9cm}}
\put(110,0){\epsfig{file=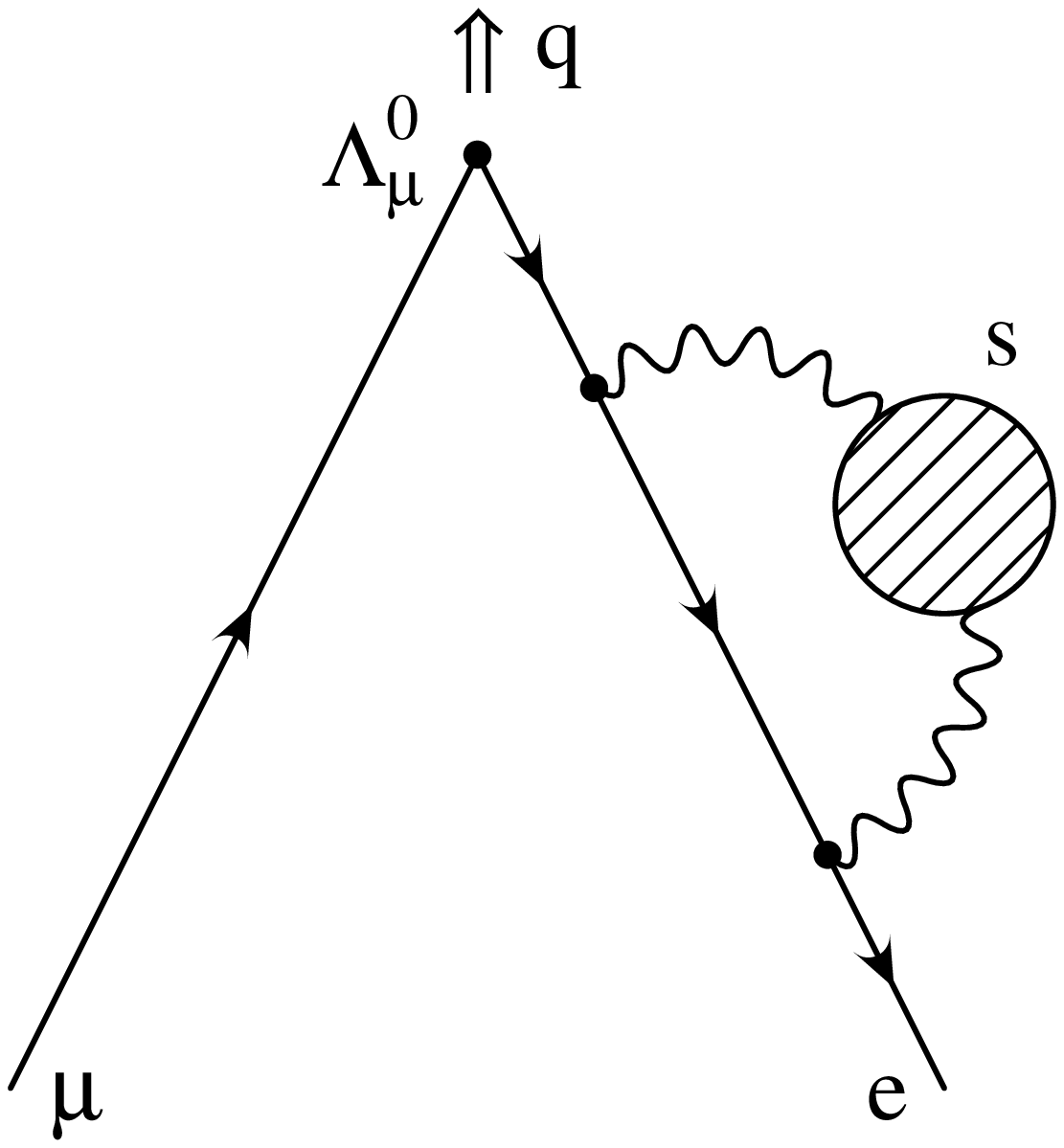,,width=4.9cm}}
\put(27,-2){a}
\put(82,-2){b}
\put(131,-2){c}
\end{picture}
\caption{\it Feynman diagrams describing a self energy insertion in the
  photonic one-loop corrections to the $\mu e$ vertex.}
\label{fig_fd}
\end{figure}

Hadronic contributions to the radiative corrections to the current
$J_{\mu}$ at order ${\cal O}(\alpha^2)$ are described by the Feynman
diagrams shown in Fig.\ \ref{fig_fd}. The hadronic vacuum polarization
\begin{equation}
\Pi^{\rm had}_{\mu \nu} (k^2) = 
\frac{-{\rm i} \Pi^{\rm had}(k^2)}{k^2 + {\rm i}0}
\left( g_{\mu \nu} - \frac{k_{\mu} k_{\nu}}{k^2} \right)
\end{equation}
is inserted in a one-loop vertex correction. The vacuum polarization can
be related to the measured cross section for $e^+e^- \rightarrow
hadrons$ with the help of a dispersion relation
\begin{equation}
\Pi^{\rm had}(k^2) = \frac{\alpha}{3\pi} \int_{s_{\rm thr}}^{\infty}
\frac{{\rm d}s}{s} R(s) \frac{k^2}{k^2 - s  + {\rm i}0} \, ,
\label{dispersion-int}
\end{equation}
where
\begin{equation}
R(s) = \frac{\sigma(s; e^+e^- \rightarrow hadrons)}%
            {\sigma(s; e^+e^- \rightarrow \mu^+ \mu^-)} 
\label{rhad}
\end{equation}
and the integration starts at the two-pion threshold, $s_{\rm thr} =
4 m_{\pi}^2$.  Therefore, the calculation corresponds to a one-loop
vertex correction with a photon of mass $\sqrt{s}$, i.e.\ using a
propagator
\begin{equation}
\frac{-{\rm i}}{k^2 - s + {\rm i}0}
\left( g_{\mu \nu} - \frac{k_{\mu}k_{\nu}}{k^2} \right) \, .
\end{equation}
The result can be written in the form
\begin{equation}
\Lambda_{\mu}(q) = \frac{\alpha}{3\pi} \int_{s_{\rm thr}}^{\infty}
\frac{{\rm d}s}{s} R(s) \widetilde{\Lambda}_{\mu}(s; q, m^2) 
\end{equation}
where the vector function $\widetilde{\Lambda}_{\mu}$ can be decomposed
into Lorentz-covariants as
\begin{equation}
\widetilde{\Lambda}_{\mu}(s; q, m^2) = 
\gamma_{\mu} \omega_L \left[ 1 + \tilde{f}(s; q^2, m^2)\right] + 
\frac{p_{\mu}+p^{\prime}_{\mu}}{m} \omega_R \tilde{g}_+(s; q^2, m^2) +
\frac{q_{\mu}}{m} \omega_R \tilde{g}_-(s; q^2, m^2) \, ,
\label{form-factor}
\end{equation}
with $\omega_{R,L} = (1 \pm \gamma_5)/2$. The calculation is
straightforward and corresponds to that of the one-loop vertex
correction in QED, with the difference that ({\it i}) the exchanged
``photon'' is massive with mass $\sqrt{s}$, ({\it ii}) the coupling is
purely left-handed, and ({\it iii}) the two fermion lines have different
masses. In fact, except for small $q^2$ the electron mass can safely be
neglected. Explicit results for the form factors $\tilde{f}$ and
$\tilde{g}_{\pm}$ are given in the appendix. After contraction with the
neutrino tensor $N_{\mu \nu}$, Eq.\ (\ref{nmunu}), only the form factors
$\tilde{f}(s; q^2, m^2)$ and $\tilde{g}_+(s; q^2, m^2)$ will remain in
the final result.

Performing the dispersion integral one obtains the form factors
\begin{equation}
\begin{array}{l}
\displaystyle
f(q^2, m^2) = \frac{\alpha}{3\pi} \int_{s_{\rm thr}}^{\infty}
\frac{{\rm d}s}{s} R(s) \tilde{f}(s; q^2, m^2) \, ,
\\[2ex] \displaystyle
g_+(q^2, m^2) = \frac{\alpha}{3\pi} \int_{s_{\rm thr}}^{\infty}
\frac{{\rm d}s}{s} R(s) \tilde{g}_+(s; q^2, m^2) \, .
\end{array}
\label{formfactors}
\end{equation}
Since we are interested in the ${\cal O}(\alpha^2)$ correction, it is
sufficient to keep only terms of first order in $f$ and $g_+$ in the
decay spectrum which then can be written in the form
\begin{equation}
\begin{array}{rcl}
\displaystyle
\frac{1}{\Gamma_0} \frac{{\rm d} \Gamma}{{\rm d} x} & = & 
2 x^2 \left[ (3 - 2x) \left(1 + 2f(x)\right) + x g_+(x) \right] \\
\displaystyle
& = & \displaystyle
\left.\frac{1}{\Gamma_0} \frac{{\rm d} \Gamma}{{\rm d} x}\right|_{\rm Born} 
(1 + r(x)) \, ,
\end{array}
\label{spectrum}
\end{equation}
with  
\begin{equation}
\Gamma_0 = \frac{G_F^2 m^5}{192 \pi^3}\, ,
\qquad 
\left. \frac{{\rm d} \Gamma}{{\rm d} x}\right|_{\rm Born} = 
2 x^2 (3 - 2x) \Gamma_0\, ,
\end{equation}
and
\begin{equation}
r(x) = 2f(x) + \frac{x}{3 - 2x} g_+(x) \, ,
\label{rx}
\end{equation}
where $f(x) \equiv f\left((1-x)m^2,m^2\right)$ 
and $g_+(x) \equiv g_+\left((1-x)m^2,m^2\right)$. 

\vspace{3mm}



\noindent
{\bf 3.} For our purpose, the function $R(s)$ describing the hadronic
cross section of $e^+e^-$ annihilation can be modeled by a combination
of experimental data and analytical results from perturbative QCD. Since
we are going to calculate a small correction, it is not necessary to
invoke the most sophisticated treatment as needed, for example, when
calculating the hadronic contribution to the fine structure constant
$\alpha(m_Z)$. At low $s < 2.5$ GeV$^2$ we use experimental data from
ALEPH parametrized in \cite{davier} or provided directly by ALEPH
\cite{aleph-tau} from a measurement of the isovector $\tau$ spectral
function. These data are complemented by the resonance contributions
from the isospin-0 light mesons $\omega$ and $\phi$. Above $s = 2.5$
GeV$^2$ we use the QCD prediction for $R(s)$ due to light quarks at
order ${\cal O}(\alpha_{\rm s})$.

Since the data in the $c\bar{c}$-channel published by various groups are
in a large part of the energy range inconsistent, we apply in this case
the QCD-based approach of analytic continuation by duality \cite{acd}.
The data region can be chosen to extend only over the sub-threshold
resonances, i.e.\ one can calculate the contribution coming from the
$c\bar{c}$-channel by a combination of data describing the $J/\Psi(1S)$
and $J/\Psi(2S)$ resonances and the prediction of perturbative QCD. We
checked that the results obtained this way are consistent with those of
the standard approach using the new BES data \cite{besii}. 



\begin{figure}[tb]
\unitlength 1mm
\vspace{-1cm}
\begin{picture}(120,110)
\put(30,0){\epsfig{file=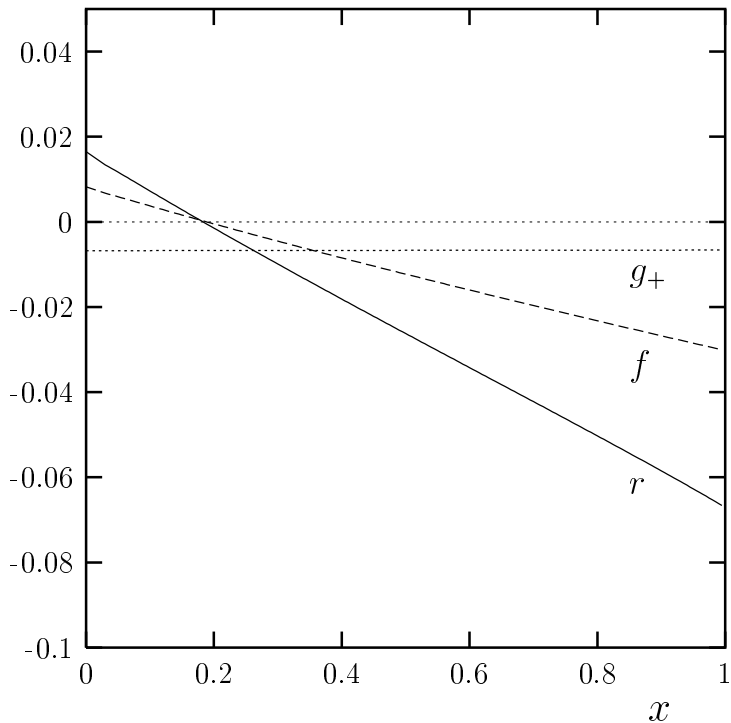,,width=10cm}}
\end{picture}
\caption{\it Results for the form factors $f$, $g_+$ and $r$ defined in
  (\ref{formfactors}), (\ref{rx}).} 
\label{fig_fgrx}
\end{figure}

The correction to the total decay rate, 
\begin{equation}
\Delta \Gamma = \int_0^1 {\rm d}x \frac{{\rm d} \Gamma}{{\rm d} x} \, ,
\end{equation} 
was calculated before in \cite{Rstuart1}.  Our result,
\begin{equation}
\Delta \Gamma_{\rm had} \simeq - 0.0421
\left(\frac{\alpha}{\pi}\right)^2  \Gamma_0 \; ,
\end{equation}
agrees perfectly with the corresponding number $-0.042$ given in
\cite{Rstuart1}.  The resulting corrections to the spectrum
(\ref{spectrum}) are shown in Fig.\ \ref{fig_fgrx}. At small $x$, the
corrections are positive, but the correction to the total decay width is
dominated by the negative values at $x \gsim 0.18$. The dependence of
the form factors on $x$ is to a very good approximation linear:
\begin{eqnarray}
f(x) & \simeq & \left(0.0071 - 0.0378 x \right) 
\left(\frac{\alpha}{\pi}\right)^2 \Gamma_0 
\, , \nonumber\\
g_+(x) & \simeq & - 0.0067 \left(\frac{\alpha}{\pi}\right)^2 \Gamma_0  
 \, , \\
r(x) & \simeq & \left(0.0148 - 0.0813 x \right) 
\left(\frac{\alpha}{\pi}\right)^2 \Gamma_0 
\, . \nonumber
\end{eqnarray}
For $g_+$, the coefficient of the term linear in $x$ is very small and
is therefore omitted.  Note that this behaviour cannot be described by a
simple redefinition of the Michel parameter. Since $G_F$ is a free
parameter in the Fermi theory, the correction to the total decay width
is not observable; it can be absorbed by a suitable redefinition of the
Fermi constant.  However, the modification of the spectrum is, in
principle, observable.

\begin{table}[bt]
\begin{center}
\begin{tabular}{rlrr}
\hline\hline
\multicolumn{4}{c}{ Contributions to $\Delta \Gamma_{\rm had}$} \\
\hline
1 & $0 < s < 0.2$ GeV$^2$                   & $- 0.00129$ & 3.1 \% \\
2 & $\omega$                                & $- 0.00223$ & 5.3 \% \\
3 & $\phi$                                  & $- 0.00264$ & 6.3 \% \\
4 & $0.2 < s < 2.5$ GeV$^2$                 & $- 0.02804$ & 66.6 \% \\
5 & $s > 2.5$ GeV$^2$                       & $- 0.00564$ & 13.4 \% \\
6 & $J/\Psi(1S)$                            & $- 0.00066$ & 1.6 \% \\
7 & $J/\Psi(2S)$                            & $- 0.00017$ & 0.4 \% \\
8 & charm, $s > 4 m_c^2$                    & $- 0.00138$ & 3.3 \% \\
9 & bottom, $s > 4 m_b^2$                   & $- 0.00003$ & 0.1 \% \\
\hline
  & Sum                                     & $- 0.04207$ & 100 \% \\
\hline\hline
\end{tabular}
\caption{Contributions to the corrections of the total decay rate}
\label{tab_Gtot}
\end{center}
\end{table}

The correction to the total decay rate can be split up into the various
contributions to the hadronic vacuum polarization, as shown in Table
\ref{tab_Gtot}. The form factor $f$ of the $\gamma_{\mu}$ term
contributes $-0.0387$, whereas the correction due to $g_+$, $-0.0033$,
is smaller by one order of magnitude. The total correction (the
contributions due to $f$ and $g_+$) is saturated to 81\,\% (80.8\,\% and
89\,\%, respectively) by the contributions from small $s$ below 2.5
GeV$^2$.  Only 5\,\% of $\Delta \Gamma_{\rm had}$ is due to charmed
states, and the bottom sector is completely negligible.

The numerical results given here take into account the ${\cal
  O}(\alpha_{\rm s})$ QCD corrections in $R(s)$. Using the leading-order
expression for $R(s)$, i.e.\ without the correction factor $(1 +
\alpha_{\rm s}/\pi)$ in the light-quark contribution at large $s$ and
the charm-quark contribution, the final result would be $-0.04149$,
i.e.\ changed by 1.4\,\%. We conclude that a more refined treatment
which would include higher orders of perturbative QCD and mass-dependent
corrections in the heavy-quark sector is not required for our purpose.

The evaluation of the dispersion integrals has been performed using
standard numerical integration routines up to a value $s_{\rm max}$ of
several hundred GeV$^2$. The contribution above this value was obtained
with the help of {\tt Maple} using the asymptotic expansion of the
form factors for large $s$ (see Eqs.\ (\ref{fasys})--(\ref{g-asys}) in
the Appendix). For intermediate values of $s$, good consistency of both
procedures has been verified.   

\vspace{3mm}

\noindent
{\bf 4.}  The same set of formulae can be used to calculate the
contributions from a $\mu^+ \mu^-$ or a $\tau^+ \tau^-$ loop insertion.
We find that the tau loop gives a very small contribution, about
$1.5\,\%$ of the one from the muon loop, in agreement with Ref.\ 
\cite{Rstuart1}. Therefore we give only results for the muon loop, where
one has to insert in Eq.\ (\ref{formfactors})
\begin{equation}
R(s) \rightarrow \left(1 + \frac{2 m^2}{s} \right) 
                 \sqrt{1 - \frac{4 m^2}{s}}
\quad {\rm and} \quad 
s_{\rm thr} \rightarrow 4 m^2\, .
\end{equation}
With this input we obtain 
\begin{equation}
\Delta \Gamma_{\rm muon} \simeq - 0.0364 \left(\frac{\alpha}{\pi}\right)^2
\Gamma_0 \; , 
\end{equation}
which perfectly agrees with the exact result given in Ref.\ 
\cite{Rstuart1}.  The results of a linear fit of the form factors for
the muon-loop insertion are:
\begin{eqnarray}
f_{\rm muon}(x) & \simeq & \left(0.0130 - 0.0414 x \right) 
\left(\frac{\alpha}{\pi}\right)^2 \Gamma_0 \, , 
\nonumber\\
g_{+,{\rm muon}}(x) & \simeq & \left( - 0.0090 + 0.0005 x \right) 
\left(\frac{\alpha}{\pi}\right)^2 \Gamma_0 \, , \\
r_{\rm muon}(x) & \simeq & \left(0.0267 - 0.0898 x \right) 
\left(\frac{\alpha}{\pi}\right)^2 \Gamma_0 \, . \nonumber
\end{eqnarray}

\vspace{3mm}



\noindent
{\bf 5.} To summarize, we found that the energy spectrum in the decay of
an unpolarized muon is corrected by a smooth function in $x$ due to
hadronic contributions at order ${\cal O}(\alpha^2)$. At both ends of
the spectrum no particularly outstanding enhancement or suppression is
observed. 

The calculations described in this paper constitute only the most
straightforward part of a full calculation which would be necessary
before the expected future high-precision data can be confronted with
theoretical predictions. This will not only be necessary for a
meaningful test of the electroweak Standard Model, but also when
searching for physics beyond the Standard Model \cite{kuno}. 


\section*{Appendix}

\newcommand{\Li}[2]{{\mbox{Li}}_{#1}\left(#2\right)}

The form factors introduced in (\ref{form-factor}) are given explicitly
in the following:
\begin{eqnarray}
\label{f}
\tilde{f}(s; q^2, m^2) & = &
- \frac{\alpha}{4\pi} \Biggl\{ 
2(m^2-q^2-s) \left[ 1+\frac{q^2 s}{(m^2-q^2)^2} \right] C_0  
\nonumber \\[1ex]
&& + \left[ \frac{2 q^2 s}{(m^2-q^2)^2}
           -\frac{2 q^2}{(m^2-q^2)}+1 \right] B_0(q^2;m^2,0)
\nonumber \\[1ex]
&& + \left[ \frac{2 q^2}{m^2-q^2}
           -\frac{s(m^2+q^2)}{(m^2-q^2)^2}\right] B_0(m^2;m^2,s)
+\left( \frac{1}{s}-\frac{1}{m^2-q^2} \right) A(s)
+ 2 \Biggl\}
\nonumber \\[1ex]
&& + \tilde{f}_{\rm SE}(s; q^2, m^2) \; ,
\\[3ex]
\label{g+}
\tilde{g}_{+}(s; q^2, m^2) &=&
- \frac{\alpha}{4\pi} \frac{m^2}{q^2}
\Biggl\{ \frac{2q^2 s}{m^2-q^2}
\left[ \frac{3q^2 s}{(m^2-q^2)^2} +2 \right] C_0
\nonumber \\[1ex]
&& -\frac{q^2}{m^2-q^2}
    \left[ \frac{6 q^2 s}{(m^2-q^2)^2} + 1 \right] B_0(q^2;m^2,0)
- \frac{q^2}{m^2-q^2}
\nonumber \\[1ex]
&& + \frac{q^2}{m^2-q^2}
    \left[ \frac{6 m^2 s}{(m^2-q^2)^2} 
          -\frac{s(4m^2-q^2)}{m^2(m^2-q^2)} +2 \right] B_0(m^2;m^2,s)
\nonumber \\[1ex]
&& + \frac{q^2}{m^2-q^2}
\left[ \frac{3}{m^2-q^2} - \frac{1}{m^2} \right] A(s)
+ \left[ \frac{1}{m^2-q^2} - \frac{1}{m^2} \right] A(m^2)
\Biggl\} \; ,
\\[3ex]
\label{g-}
\tilde{g}_{-}(s; q^2, m^2) &=&
\frac{\alpha}{4\pi} \frac{m^2}{q^2}
\Biggl\{ \frac{2 q^2 s}{m^2-q^2}
\left[ \frac{3 q^2 s}{(m^2-q^2)^2} + \frac{2 s}{m^2-q^2} +2 \right] C_0
- \frac{q^2}{m^2-q^2}
\nonumber \\[1ex]
&& +\frac{1}{m^2-q^2}
\left[- \frac{6 m^2 q^2 s}{(m^2-q^2)^2}
      + \frac{2 q^2 s}{m^2-q^2} +2 m^2 - 3 q^2 \right]
B_0(q^2;m^2,0)
\nonumber \\[1ex]
&& + \frac{q^2}{m^2-q^2}
\left[ \frac{ 6 m^2 s}{(m^2-q^2)^2}
       +\frac{s}{m^2-q^2} + \frac{s}{m^2} + 2 \right]
B_0(m^2;m^2,s)  
\nonumber \\[1ex]
&& + \frac{q^2}{m^2-q^2}
\left( \frac{1}{m^2} + \frac{3}{m^2-q^2} \right) A(s)
+ \left( \frac{1}{m^2} + \frac{1}{m^2-q^2} \right) A(m^2)
\Biggl\} \; ,
\end{eqnarray}
where $C_0 = C_0(m^2,0,q^2;m^2,s,0)$ is the three-point integral (cf.\ 
Fig.\ \ref{fig_fd}a), defined in Eq.~(\ref{C0}) below. $B_0$ and $A$
denote the tadpole and two-point integrals \cite{tHV'79,PV},
respectively (see Eqs.~(\ref{A})--(\ref{B2}) below).

Self-energy diagrams (cf.\ Fig.\ \ref{fig_fd}b, c) contribute to the
coefficient of $\gamma_{\mu}$ only and are given by
\begin{eqnarray}
\tilde{f}_{\rm SE}(s; q^2, m^2) & = &
\frac{\alpha}{8\pi} \Biggl\{
2(s + 2m^2) 
\left.\frac{\partial}{\partial p^2} B_0(p^2;m^2,s)\right|_{p^2=m^2}
- \frac{s}{m^2} B_0(m^2;m^2,s)
\nonumber \\
&& - \frac{s+m^2}{s m^2} A(s) + \frac{1}{m^2} A(m^2) + 1
+ \frac{3}{2} \Biggl\} ,
\label{fse}
\end{eqnarray}
where the last term, ${\textstyle{3\over2}}$, comes from the self energy
on the massless (electron) leg. Using recurrence relations~\cite{ibp}
(see also Appendix~A of~\cite{BDS}), the derivative in (\ref{fse}) can
be represented as
\begin{eqnarray}
\left.\frac{\partial}{\partial p^2}
B_0(p^2;m^2,s)\right|_{p^2=m^2}
&=& \frac{1}{m^2 (s-4m^2)}
\Biggl[ -(s-3m^2) B_0(m^2;m^2,s) 
\nonumber \\
&& + A(m^2) - \left(1 - \frac{2m^2}{s} \right) A(s) + m^2 \Biggl] .
\end{eqnarray}
The required tadpole and two-point integrals are~\cite{tHV'79,PV}
\begin{eqnarray}
A(m^2) & = & m^2 \left[ - \Delta  - 1 
+ \ln \frac{m^2}{\mu_{\rm DR}^2} \right],
\label{A}
\\
B_0(m^2;m^2,s) & = & \Delta + 2 - \ln \frac{m^2}{\mu_{\rm DR}^2} 
- \frac{s}{2m^2} \ln \frac{s}{m^2}
+ \frac{s}{2m^2} \beta_s
\ln\left(\frac{1+\beta_s}{1-\beta_s}\right),
\label{B1}
\\
B_0(q^2;m^2,0) & = & \Delta + 2 -
\ln \frac{m^2}{\mu_{\rm DR}^2}
+ \frac{m^2-q^2}{q^2} \ln\left(\frac{m^2-q^2}{m^2}\right) ,
\label{B2}
\end{eqnarray}
where 
\begin{equation}
\beta_s \equiv \sqrt{1-\frac{4m^2}{s}} \; ,
\end{equation}
and $\mu_{\rm DR}$ is the scale parameter of dimensional regularization.
In Eqs.~(\ref{A})--(\ref{B2}), terms containing $\Delta = 1/\varepsilon
- \ln\pi - \gamma_{\rm E}$ represent the ultraviolet singularities which
cancel in the final results (\ref{f})--(\ref{g-}).

Finally, we need the three-point scalar function $C_0$ \cite{tHV'79} for
positive values of $q^2$. In this case it can be written in the
following form: 
\begin{eqnarray}
\label{C0}
C_0(m^2,0,q^2; m^2,s,0) &=&
\frac{1}{m^2-q^2} \Biggl\{
\Li{2}{1-\frac{m^2}{q^2}}
+ \Li{2}{-\frac{(m^2-q^2)^2}{s q^2} }
\nonumber \\[1ex]
&& - \mbox{Li}_2\left[\frac{1}{2} \left(1-\frac{m^2}{q^2}\right)
         \left( 1+ \beta_s \right) \right]
- \mbox{Li}_2\left[\frac{1}{2} \left(1-\frac{m^2}{q^2}\right)
         \left( 1- \beta_s \right) \right]
\nonumber \\[1ex]
&& + \frac{1}{2} \ln\left[ \frac{(m^2-q^2)^2}{s m^2} \right]
          \ln\left[ 1+\frac{(m^2-q^2)^2}{s q^2} \right]
- \frac{1}{2} \ln\frac{m^2}{q^2} \ln\frac{m^2}{s}
\nonumber \\[1ex]
&& + \frac{1}{2} \ln\left[ 
\frac{m^2+q^2+(q^2-m^2)\beta_s}{m^2+q^2-(q^2-m^2)\beta_s} \right]
\ln\left( \frac{1+\beta_s}{1-\beta_s} \right)
\Biggl\} \; .
\end{eqnarray}

For convenience, we also give asymptotic expansions of the form factors
(\ref{f})--(\ref{g-}) valid for large $s$, 
\begin{eqnarray}
\frac{4\pi}{\alpha} \tilde{f} \!\! & = & \!\!
- \frac{1}{3 s (q^2)^2}
\Biggl\{ (m^2-q^2)^2 (m^2+2q^2) \ln\left( \frac{m^2-q^2}{m^2} \right)
+ (q^2)^2 (3 m^2-2 q^2) \ln\frac{s}{m^2}
\nonumber \\
&& + q^2  \left[ m^4- \frac{9}{2} m^2 q^2- \frac{11}{3} (q^2)^2 \right]
\Biggl\}
\nonumber \\
&& + \frac{1}{6 s^2 (q^2)^3}
\Biggl\{ (m^2\!-\!q^2)^4 (m^2\!+\!q^2) \ln\left( \frac{m^2\!-\!q^2}{m^2}
\right)
\!- (q^2)^3 \left[ 35 m^4\!-\!3 m^2 q^2\!+\!(q^2)^2 \right] \ln\frac{s}{m^2}
\nonumber \\  
&& + q^2
\left[ m^8-\frac{5}{2} m^6 q^2 + \frac{697}{12} m^4 (q^2)^2
      + \frac{11}{4} m^2 (q^2)^3 - \frac{13}{12} (q^2)^4 \right]
\Biggl\} + {\cal O}(s^{-3}) \; ,
\label{fasys}
\end{eqnarray}
\begin{eqnarray}
\frac{4\pi}{\alpha}\tilde{g}_{+} \!\! &=& \!\!
- \frac{m^2}{3 s}
+ \frac{m^2}{6 s^2 (q^2)^3}
\Biggl\{ (m^2-q^2)^4 \ln\left( \frac{m^2-q^2}{m^2} \right)
+(q^2)^3 (4 m^2-q^2) \ln\frac{s}{m^2}
\nonumber \\[1ex]
&& +q^2 \left[ m^6-\frac{7}{2} m^4 q^2 - \frac{8}{3} m^2 (q^2)^2
 +\frac{5}{12} (q^2)^3 \right] 
\Biggl\}
+ {\cal O}(s^{-3}) \; ,
\label{g+asys}
\\[3ex]
\frac{4\pi}{\alpha}\tilde{g}_{-}  \!\! &=& \!\!
\frac{4 m^2}{3 s (q^2)^3)}
\Biggl\{ (m^2\!-\!q^2)^3 \ln\left( \frac{m^2\!-\!q^2}{m^2} \right)
+ (q^2)^3 \ln\frac{s}{m^2}
+ q^2 \left[ m^4 \!-\! \frac{5}{2} m^2 q^2 
\!+\! \frac{7}{12} (q^2)^2 \right]
\Biggl\}
\nonumber \\[1ex]
&&  - \frac{m^2}{6 s^2 (q^2)^4}
\Biggl\{ (m^2-q^2)^4 (6 m^2-5q^2) \ln\left( \frac{m^2-q^2}{m^2} \right)
         -(q^2)^4 (26 m^2-5 q^2) \ln\frac{s}{m^2}
\nonumber \\[1ex]
&& +q^2 \left[ 6 m^8 - 26 m^6 q^2 + \frac{87}{2} m^4 (q^2)^2
- \frac{25}{6} m^2 (q^2)^3 + \frac{23}{12} (q^2)^4 \right]
\Biggl\} + {\cal O}(s^{-3}) \; .
\label{g-asys}
\end{eqnarray}
These expressions turned out to be useful for the numerical evaluation
of the dispersion integrals in the large-$s$ region.

The asymptotic values for $x=0$ are as follows:
\begin{eqnarray*}
\left.
\frac{4\pi}{\alpha}\; \tilde{f}\right|_{x=0} \!\! & = & \!\! 
- \frac{1}{12 m^4} 
  \Biggl\{ (s\!+\!2m^2) (7 s\!-\!22 m^2)\frac{1}{\beta_s}
         \ln\left( \frac{1\!+\!\beta_s}{1\!-\!\beta_s} \right)
          +(18 m^4\!-\!6 m^2 s \!-\! 7 s^2)\ln\frac{s}{m^2}
\\
&& +m^2 (33 m^2+14 s) \Biggl\},
\\[3ex]
\left.
\frac{4\pi}{\alpha} \tilde{g}_{+}\right|_{x=0} \!\! & = & \!\!
\frac{1}{4 m^4}
\Biggl\{ s (2 m^2\!-\!s) \beta_s  
\ln\left( \frac{1\!+\!\beta_s}{1\!-\!\beta_s} \right)
         +(2 m^4\!-\!4 m^2 s \!+\!s^2) \ln\frac{s}{m^2}
+m^2 (5 m^2 \!-\! 2 s) \Biggl\},
\\[3ex]
\left.
\frac{4\pi}{\alpha} \tilde{g}_{-}\right|_{x=0} \!\! & = & \!\!
\frac{1}{12 m^4}
\Biggl\{ s (2 m^2-5 s) \beta_s
\ln\left( \frac{1+\beta_s}{1-\beta_s} \right)
         -(6 m^4+12 m^2 s -5 s^2) \ln\frac{s}{m^2}
\\ &&
+m^2 (9 m^2-10 s) \Biggl\} .
\end{eqnarray*}
Note that there are no contributions of the order ${\cal O}(x\ln x)$.


\end{document}